\documentclass[trackchanges]{aastex701}
\usepackage{afterpage}
\usepackage{newtxtext,newtxmath}
\usepackage{array}
\usepackage{multirow}
\usepackage[flushleft]{threeparttable}
\usepackage{caption}

\usepackage[T1]{fontenc}
\usepackage{makecell}
\usepackage[version=4]{mhchem} 
\DeclareRobustCommand{\VAN}[3]{#2}
\let\VANthebibliography\thebibliography
\def\thebibliography{\DeclareRobustCommand{\VAN}[3]{##3}\VANthebibliography}
\usepackage{graphicx}	
\usepackage{amsmath}	

\usepackage[flushleft]{threeparttable}

\begin{document}

\title{Analysis of Eruptive Prominence Plasma Parameters' Effects on the
\ion{He}{2} 304~\AA\ Line with Solar Orbiter EUI Observations}

\author[orcid=0009-0003-2989-7421,gname=Yong, sname='Zhang']{Yong Zhang}
\affiliation{SUPA School of Physics and Astronomy, University of Glasgow, Glasgow G12 8QQ UK}
\email[show]{y.zhang.9@research.gla.ac.uk}  

\author[orcid=0000-0002-4638-157X,gname=Nicolas, sname=Labrosse]{Nicolas Labrosse} 
\affiliation{SUPA School of Physics and Astronomy, University of Glasgow, Glasgow G12 8QQ UK}
\email[show]{nicolas.labrosse@glasgow.ac.uk}

\author[orcid=0000-0002-9242-2643,gname=Sargam,sname=Mulay]{Sargam M. Mulay}
\affiliation{SUPA School of Physics and Astronomy, University of Glasgow, Glasgow G12 8QQ UK}
\email[show]{Sargam.Mulay@glasgow.ac.uk}






\begin{abstract}

An observation of a large prominence on the solar limb took place on February 15, 2022, by the Extreme Ultraviolet Imager (EUI) on board Solar Orbiter. We aim to determine the range of physical parameters of this prominence, such as temperature, radial velocity, and altitude, and examine how these parameters affect the formation of the 304~\AA\ line of \ion{He}{2}, especially how collisional excitation and radiative processes contribute to line formation.

After constraining these parameters, we generate 200 random models and compute the \ion{He}{2} 304~\AA\ line profile. We present these results using parallel coordinate plots to explore how these parameters affect the results. This allows us to infer the key physical parameters that impact the formation of the \ion{He}{2} 304~\AA\ line. 

This study demonstrates that column mass and the steepness of the temperature profile are key factors in the formation of the \ion{He}{2} 304~\AA\ line during the solar prominence eruption on February 15, 2022. Radiative processes remain dominant in the formation of the \ion{He}{2} 304~\AA\ line. These insights provide a foundation for future research and comparative studies.

\end{abstract}

\keywords{\uat{Solar prominences}{1519} --- \uat{Solar coronal lines}{2038}}


\section{Introduction}\label{sec:intro}

The study by \cite{mierla_2022_prominence}  on prominence kinematics and radiative properties based on an observation obtained on February 15, 2022, by the Extreme Ultraviolet Imager (EUI; \citealt{rochus2020solar}) on board Solar Orbiter \cite{muller2020solar} provides a foundation for focusing on line formation and determining whether the observed \ion{He}{2} 304~\AA\  emission primarily originates from resonant scattering or collisional excitation. \cite{mierla_2022_prominence} conjecture that the 304~\AA\ line in this eruptive prominence is due to collisional excitation rather than to resonant scattering, but a concrete conclusion is not reached. This is the key question we want to explore in this paper. In this study, we analyze the same prominence eruption event. Specifically, we investigated the mechanisms behind the formation of the \ion{He}{2} 304~\AA\ line during the eruption and how plasma diagnostics could enhance our understanding of the dynamics involved. 

Previous studies have laid the essential groundwork for understanding the formation of the \ion{He}{2} 304~\AA\ line in solar prominences. \cite{labrosse2001formation} presented non-local thermodynamic equilibrium (non-LTE) modelling of the Helium spectrum in isothermal isobaric static prominences, demonstrating how line intensities are influenced by variations in temperature, pressure, and microturbulence. This was then extended in \cite{labrosse2004} to show how the introduction of a prominence-to-corona transition region impacts the energy level populations within the helium atom via the coupling between statistical equilibrium and radiative transfer equations.  \cite{labrosse2007effect} advanced this work by examining the Doppler dimming effect in moving prominences, highlighting how radial mass motion influences the \ion{He}{2} 304~\AA\ line. \cite{gouttebroze2009radiative} extended these models with a cylindrical-thread approach, incorporating two-dimensional structural variations that provide a refined understanding of intensity distributions within prominences.

\cite{wang1998comparison} provided early observational evidence of prominences exhibiting distinct morphological differences in H$\alpha$ and \ion{He}{2} 304~\AA, challenging the assumption that these lines trace similar structures. \cite{schiralli2014comparison} has revealed important differences between H$\alpha$ and \ion{He}{2} 304~\AA\ observations, showing that while H$\alpha$ provides detail in cooler prominence cores, 304~\AA\ line emission better captures the structure, including hotter transition regions. \cite{labrosse2011euv} further examined \ion{He}{2} 256.32~\AA\ with Hinode/EIS data, confirming its formation predominantly via resonant scattering in the prominence-to-corona transition region (PCTR). \cite{labrosse2012plasma} applied diagnostics of Doppler effects to prominence eruptions observed at 304~\AA, confirming that Doppler shifts provide critical information on prominence plasma conditions during eruptions. \cite{labrosse2016radiative} introduced multi-thread models that simulate the complex interactions within prominence fine structures, demonstrating how the multi-thread structure affects the helium line shape and intensities, and temperature gradients strongly impact the Helium spectrum. Together, these studies offer a framework for interpreting \ion{He}{2} line formation. 

In this work, we build upon these insights to explore the \ion{He}{2} 304~\AA\ line formation in eruptive prominences and advance prominence plasma diagnostics. We present the observations in Section~\ref{sect2:observations}. The non-LTE models are described in Section~\ref{Sect3:non-lte-model}.
A filter-ratio analysis of the STEREO EUVI-A data is presented in Section~\ref{Sect4:filter-ratio} to explore the temperature distribution of the eruptive event. Kinematic analysis, including position and velocity calculations of distinct features, is performed using EUI/FSI data. We explore how different parameters affect the formation of the 304~\AA\ line of \ion{He}{2}, especially how collisional excitation and radiative processes contribute to line formation.
These results are presented in Section~\ref{Sec: compare}. The conclusion of this work is presented in Section~\ref{Sect6:conclusion}.

\section{Observations} 
\label{sect2:observations}
A prominence eruption was observed in the \ion{He}{2} 304~\AA\ passband by EUI/FSI telescope aboard Solar Orbiter on February 15-16, 2022. This is the first observed eruption in the \ion{He}{2} 304~\AA\ emission that is above 6~R$_{{\ensuremath{\odot}}}$ \citep{mierla_2022_prominence}. We also have EUVI-A observation in 171~\AA, 195~\AA\ and 284~\AA\ channels. 
\subsection{Solar Terrestrial Relations Observatory (STEREO)} 
\label{sect2.1:STEREO-obs}

The Solar Terrestrial Relations Observatory (STEREO) spacecraft, launched in 2006, aims to study Coronal Mass Ejections (CMEs) and solar wind dynamics \citep{kaiser2008stereo}. Two spacecraft orbit the Sun at approximately 1~AU, with one spacecraft leading Earth (STEREO-A) and the other trailing behind (STEREO-B). Equipped with various instruments, including coronagraphs and particle sensors, STEREO provides crucial data to understand space weather and predict CME impacts on Earth. STEREO-B has been out of contact since Oct 1, 2014. However, the remaining STEREO-A still provides a complementary view of solar structures with a different line of sight from Earth-based observations. 

The Extreme Ultraviolet Imager (EUVI) \citep{wuelser2004euvi} is a component of the Sun-Earth Connection Coronal and Heliospheric Investigations (SECCHI) instrument suite developed for NASA's STEREO mission \citep{howard2008sun}. It focuses on the initiation and early evolution of CMEs. The EUVI features 2048 x 2048 pixel detectors that provide a field of view extending to 1.7 solar radii and record images in four EUV spectral channels, 171, 195, 284 and 304~{\AA} covering a temperature range from 0.1 to 20~MK. We used images from the EUVI-A 171 and 195~{\AA} channels to study the prominence eruption that occurred on February 15, 2022. The EUVI level~0.5 data were downloaded and converted to level~1 using the solarsoft (SSW; \citealt{1998_Freeland_solarsoft}) routine\footnote{\url{https://stereo-ssc.nascom.nasa.gov/publications/CMAD/secchi/STEREO\_SECCHI\_Prep\_CMAD\_20211206.pdf}}  \texttt{sechhi\_prep.pro}. We obtained the EUVI images with the intensity units of DN and normalised them using the exposure time (units = DN/s/pix). These images were used further for temperature diagnostics using the filter ratio method (see section~\ref{Sect4:filter-ratio}).

\subsection{Solar Orbiter} 
\label{sect2.2:solar-orbiter-obs}

\begin{figure}
    \centering
\includegraphics[trim=0cm 3cm 0cm 3cm, width=0.6\textwidth]{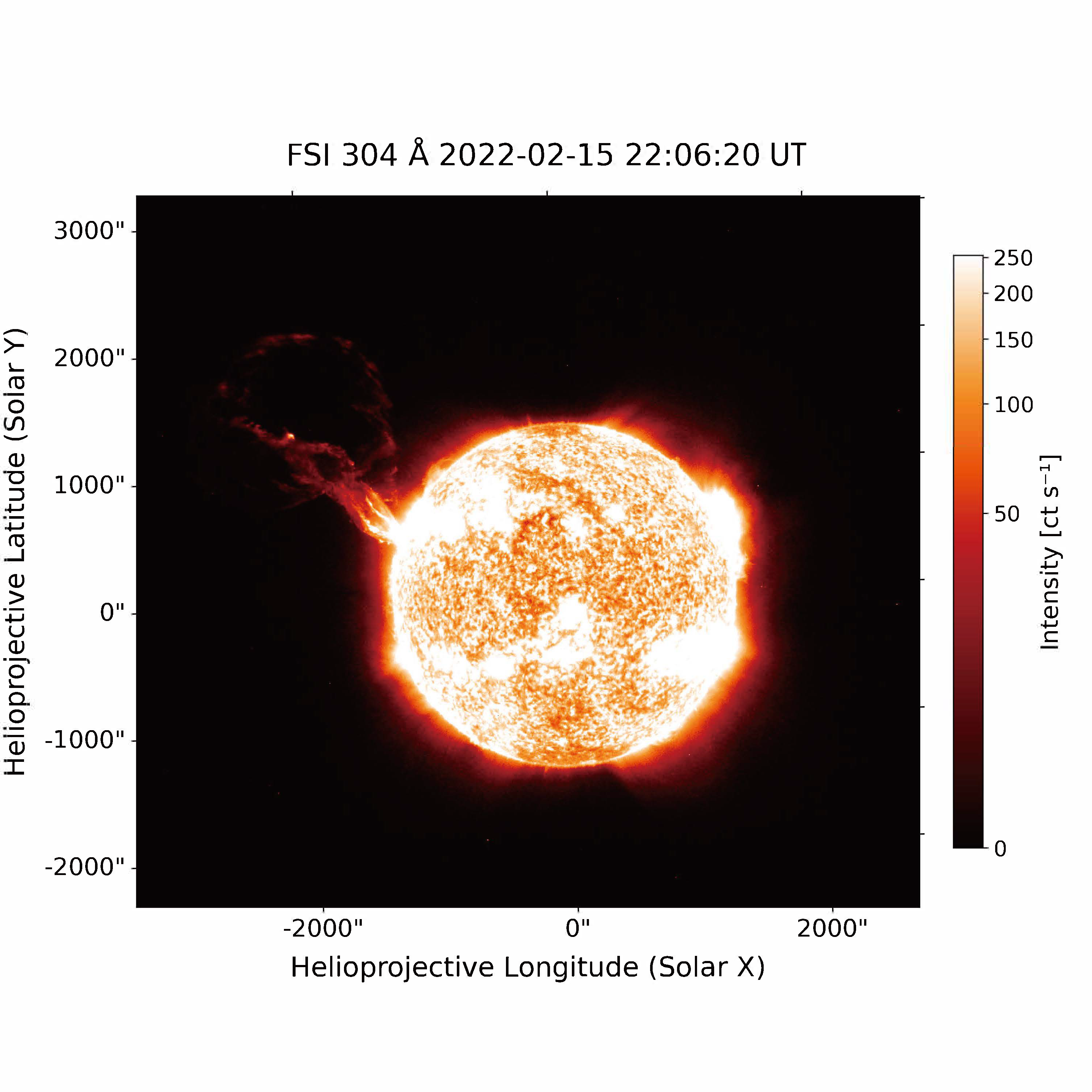}
    \caption{The full disk image of the Sun with the prominence eruption observed on February 15, 2022, using EUI/FSI 304~\AA channel.}
    \label{fig:euiobs}
\end{figure}

Solar Orbiter \citep{muller2020solar} is a Sun-observing satellite developed by the European Space Agency (ESA). The highly elliptical orbit allows it to go close to the Sun. Since 2025, the orbit will be inclined with respect to the ecliptic, allowing for the view of the poles. The Extreme Ultraviolet Imager (EUI; \citealt{rochus2020solar}) is on board the Solar Orbiter. EUI captures images of the Sun's atmosphere in the extreme ultraviolet range, providing information about the Sun's corona, solar flares, and other phenomena.   Full-Sun Imager (FSI) is a part of the EUI instrument suite on board the Solar Orbiter spacecraft.  FSI is designed to observe the full solar disk in two EUV channels: 174~\AA\ and 304~\AA. The observations at various wavelengths enable the acquisition of detailed visual data of the sun's entire surface, thereby facilitating scientific analysis of various solar layers and phenomena. Figure~\ref{fig:euiobs} shows a full disk image of the Sun along with a prominence eruption observed in the 304~\AA\ channel of EUI/FSI on February 15, 2022. We adjusted the intensity scale in this image to reveal the faint EUV emission from the prominence eruption. This adjustment made the disk emission appear brighter.

\label{sect2.3:EUI_channels}

EUI observes the sun in multiple channels. Of particular interest to this project is the 304~\AA\ channel. The 304~\AA\ channel is sensitive to cooler plasma in the transition region between the solar chromosphere and corona, with emission lines originating from singly-ionised helium (\ion{He}{2}). The 304~\AA\ line has demonstrated high sensitivity to the radial movement of the solar prominence. This is due to its formation primarily occurring through the scattering of incident radiation from the Sun \citep{labrosse2012plasma}. It provides information on the dynamics of prominences, filaments, and other features on the solar disk \citep{mierla_2022_prominence}. Such observations help us better understand the mechanisms responsible for heating the corona, accelerating the solar wind, and driving space weather phenomena. 

\section{Non-LTE modelling} 
\label{Sect3:non-lte-model}

A non-LTE model is used to describe the physical properties of plasma in the solar atmosphere. The non-LTE approach is important because the solar prominence plasma is far from local thermodynamic equilibrium (LTE). We obtain different spectral profiles using this modelling approach. To obtain the emergent intensities in the \ion{He}{2} 304~\AA\ line, the radiative transfer and statistical equilibrium equations for Hydrogen and Helium are solved. This is done assuming that the plasma is in ionisation equilibrium and steady state, meaning that the sum of the excitation and de-excitation processes acting on an energy level of the ion keeps its population constant in time. For this prominence eruption event, we use a 1D slab standing vertically above the solar surface to represent the eruptive prominence and consider the effect of the prominence-to-corona transition region. The prominence moves with some radial velocity, which is a model input. We also need to input the temperature and pressure in the slab center and the boundary with the corona, the shape of the temperature gradient in the PCTR region, the total column mass, the microturbulent velocity, and the altitude of the slab \citep{labrosse2012plasma}. The temperature and pressure profiles in the models are from \cite{anzer1999energy}. The mathematical expressions for the temperature and pressure profiles are given by equations \ref{T} and \ref{p}, respectively \citep{anzer1999energy}. 

\begin{equation}
\begin{split}
{ T ( m ) = T _ { \text {cen} } + ( T _ { \text {tr} } - T _ { \text {cen} } ) ( 1 - 4 \frac { m } { M } ( 1 - \frac { m } { M } ) ) ^ { \gamma } } 
\end{split}
\label{T}
\end{equation}
\begin{equation}
\begin{split}
{ p ( m ) = 4 p _ { c } \frac { m } { M } ( 1 - \frac { m } { M } ) + p _ { \text {tr} }  } 
\end{split}
\label{p}
\end{equation}

Here, $p_0$ is the pressure at the outer boundary, and $p_c$ is the difference between the pressure at the center and at the outer boundary.  $p_{cen}=p_c+p_0$. The column mass $m$ ranges from $m=0$ to $m=M$ (from one surface to the other surface). $\gamma$ determines the temperature gradient of the PCTR region. 
$T_{\text {cen}}$ is the central temperature and $T_{\text {tr}}$ is the boundary transition-region temperature. The column mass $m$ ranges from $m=0$ to $m=M$ (from one surface to the opposite surface). The computational methodology is detailed in \citet{labrosse2007effect, labrosse2008diagnostics}. We require estimates for the ranges of input parameters prior to inputting them into the model. Subsequently, the model can be used to investigate the formation of different spectral line profiles.

\section{Filter-ratio analysis}
\label{Sect4:filter-ratio}


Plasma diagnostic techniques, such as the filter-ratio method \citep{del2011sdo, 2017_Mulay_filter_ratio_jets}, have been used to study plasma temperature during various dynamic events in the solar atmosphere. Particularly, broadband filters from the Atmospheric Imaging Assembly (AIA; \citealt{2012_Lemen_AIA_SDO}) instrument onboard Solar Dynamics Observatory (SDO) and X-ray Telescope (XRT; \citealt{2007_Golub_XRT}) onboard Hinode satellite \citep{2007_Kosugi_Hinode} have been extensively exploited. This method involves studying emitted radiation using two broadband filters from the spacecraft and comparing the intensity ratios recorded at each pixel with the corresponding ratios of the temperature response function of those filters.  

\begin{figure*}
    \centering
    \includegraphics[trim=0cm 2cm 0cm 3cm, width=1.0\textwidth]{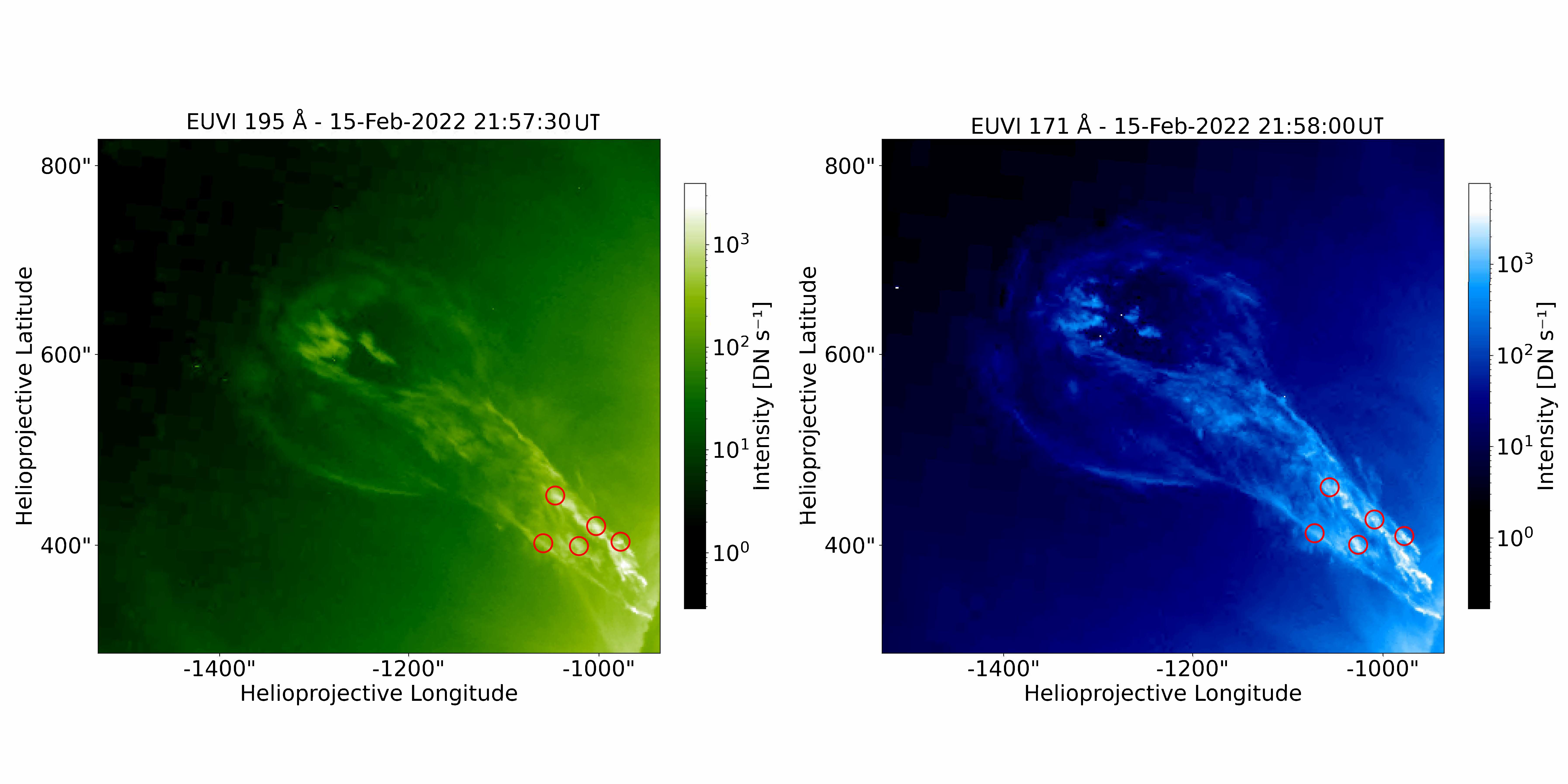}
    \caption{The prominence observed by STEREO EUVI-A in the 171~\AA\ and 195~\AA\ channels. The five red circles are the five features we tracked. }
    \label{fig:171_195}
\end{figure*}

For the February 15, 2022, prominence eruption event, the field of view of AIA is relatively small. It is not possible to track the prominence eruption up to a high altitude (6~R$_{{\ensuremath{\odot}}}$). Moreover, the prominence is behind the limb from the viewpoint of AIA, and we could not use these images to study the temperature using the filter ratio technique. Instead, we focused on near simultaneous images from the 171~\AA\ and 195~\AA\ channels of STEREO EUVI-A, where the prominence eruption was nicely captured. The images are shown in Fig.~\ref{fig:171_195}. We studied the evolution of the prominence at various heights and used the observed intensity (in units of DN/s/pix) recorded in these filters to analyse the temperature of the erupting prominence.

We make several assumptions when applying the filter-ratio analysis to the EUVI-A 171~\AA\ and 195~\AA\ channels. First, we assume that the plasma is optically thin in the range of wavelengths used in this analysis, so that the emission from the plasma at a given wavelength should not be significantly absorbed or scattered by the plasma itself, allowing the observed intensity to be directly related to the amount of emitting material along the line of sight. This assumption aligns with the typical characteristics of PCTR regions. Second, to simplify the analysis, we assume an isothermal temperature distribution for prominence structures along the line of sight. Third, we assume a constant temperature over time within each feature during the brief observation period (21:57:30 $\sim$ 21:58:00~UT) because of the small differences in the observation times between the 171~\AA\ and 195~\AA\ channels. This is reasonable since temperature variations are generally small over short timescales. Finally, we acknowledge minor alignment variances due to manual tracking, though careful frame-by-frame adjustments mitigate this risk. 

\begin{figure}
    \centering
    \includegraphics[width=0.45\textwidth]{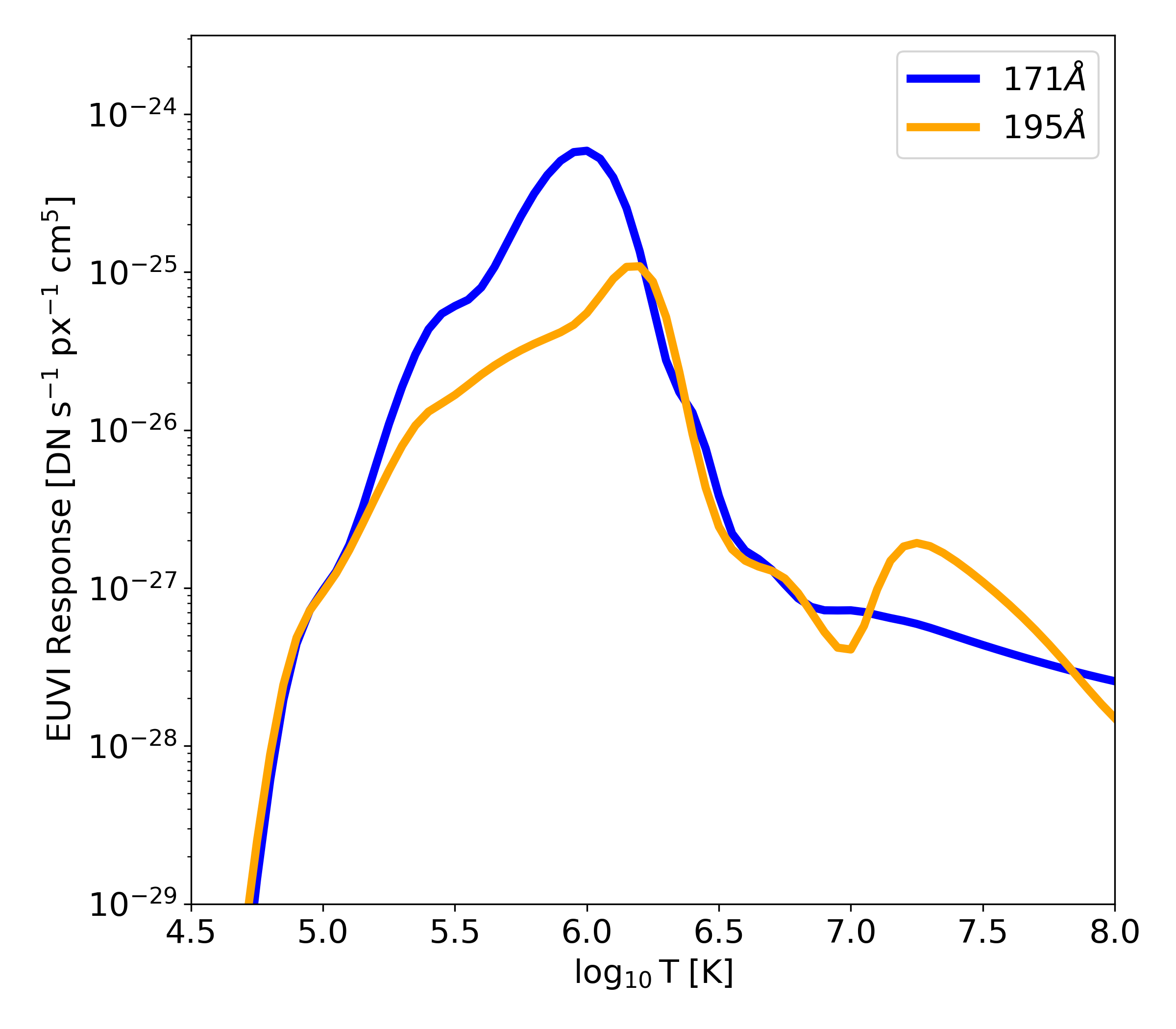}
    \includegraphics[width=0.45\textwidth]{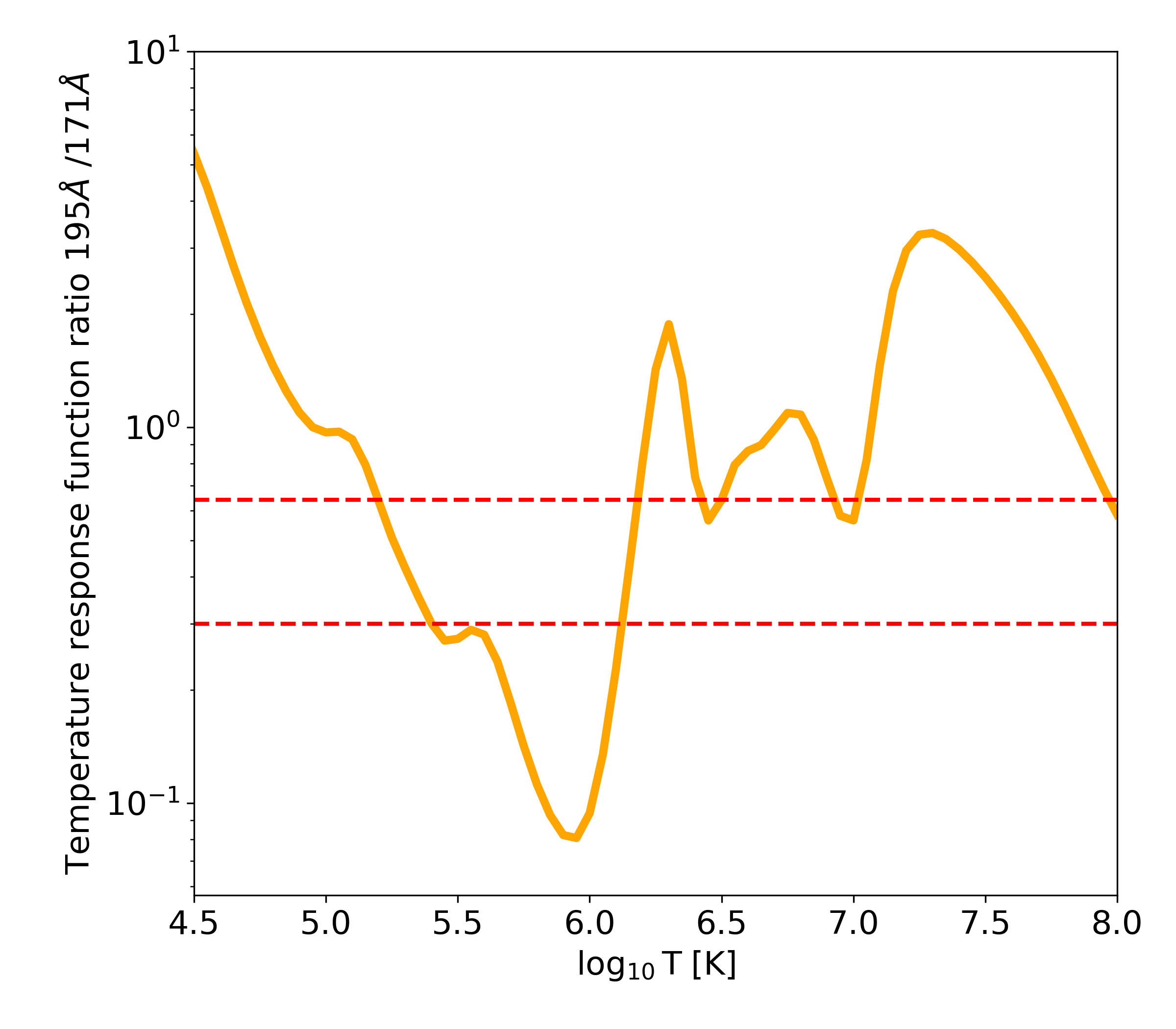}
    \caption{Left panel: The STEREO EUVI-A temperature response functions for the 171~\AA, 195~\AA\ and 284~\AA\ channels. They are calculated using the effective area from SSW and the CHIANTI atomic database version 11.0.2. Right panel: The ratio of temperature response functions for the EUVI 195~\AA\ and 171~\AA\ channels. The two red dashed lines are the minimum and maximum ratios in Table \ref{table:171_195}.}
    \label{fig:Response}
\end{figure}

The temperature response function (in DN s$^{-1}$ pix$^{-1}$ cm$^{5}$ units) for the 171~\AA\ and 195~\AA\ channels of EUVI-A are calculated using the CHIANTI atomic database version 11.0.2 \citep{dufresne2024chianti}, electron number density, \textit{N}$_{\rm e}$ of 1$\times$10$^{10}$~cm$^{-3}$, and coronal abundances. We used the effective area\footnote{\url{https://stereo-ssc.nascom.nasa.gov/publications/CMAD/secchi/STEREO\_SECCHI\_EUVI\_CMAD\_20211206.pdf}} for these filters from SSW. We followed a similar procedure (see Appendix of \citealt{del2011sdo}) used to calculate temperature responses for the AIA filters to calculate those for the STEREO EUVI filters. Figure~\ref{fig:Response} (left panel) shows the temperature response function for the 171~\AA, 195~\AA\ and 284~\AA\ channels, and the ratio of the temperature responses for the 195/171 filters is shown in the right panel.

The temperature responses of these three filters show multiple peaks across a wide temperature range, indicating their multithermal nature. The ratio of the responses is not a single-valued function (like XRT filter ratio - see Fig.~A3 in \citealt{2017_Mulay_filter_ratio_jets}), so we carefully considered the limitations in interpreting the temperature information from this filter ratio analysis.


The peaks in the filter temperature responses indicate that the photons in the EUVI 171~\AA\ and 195~\AA\ channels are emitted mostly from the PCTR region and by the surrounding coronal plasma, which should be optically thin. These filters are suitable for the filter-ratio analysis of the erupting prominence region. We chose five features along the prominence, shown with red circles in Fig.~\ref{fig:171_195}. We manually identified and tracked these features from 22:00 to 22:10~UT. We also assumed that the temperatures of the features we track are constant across the domain and over time to apply the filter-ratio method.



\begin{table}
\hspace*{-1.75cm} 
\small
  \begin{threeparttable}
    \caption{The observed intensity ratios of 195~{\AA}/171~{\AA} filters for five features along the eruptive prominence.}
     \begin{tabular}{c c c c c c c}
\hline
Feature  &\multicolumn{2}{c}{Intensity of 171~{\AA} [DN/s]} &\multicolumn{2}{c}{Intensity of 195~{\AA} [DN/s]} &195/171~{\AA} intensity ratio &log temperature [K] (MK)\\
 &before BS &after BS &before BS &after BS &after BS & \\
\hline
1 & 1650 & 1493 & 1098 & 961 & 0.64  & \makecell{5.20 (0.16), 6.18 (1.51), 6.43 (2.67), 6.50 (3.15),\\ 6.93 (8.53), 7.02 (10.35), 7.97 (93.75)}\\ 
2 & 904 & 736 & 517 & 348 & 0.47 & 5.27 (0.19), 6.16 (1.43)\\ 
3 & 3024 & 2817 & 1755 & 1532 & 0.54 & 5.24 (0.17), 6.17 (1.47)\\
4 & 2462 & 2215 & 896 & 661 & 0.30  & 5.40 (0.25), 6.12 (1.31)\\ 
5 & 3297 & 2980 & 1730 & 1340 & 0.45 & 5.29 (0.19), 6.15 (1.43)\\

\hline
\end{tabular} 
    \begin{tablenotes}
      \small
      \item \textbf{Notes -} Intensity values are given before and after background subtraction (BS). We choose intensity values at the center of circles in Figure~\ref{fig:171_195}. The temperature values are for the features shown in Figure~\ref{fig:171_195}. The features 1-5 are labelled in decreasing order of distance from the limb.
    \end{tablenotes}\label{table:171_195}
  \end{threeparttable}
\end{table}

        

\afterpage{
\begin{figure*}
    \centering
\includegraphics[width=1.05\textwidth]{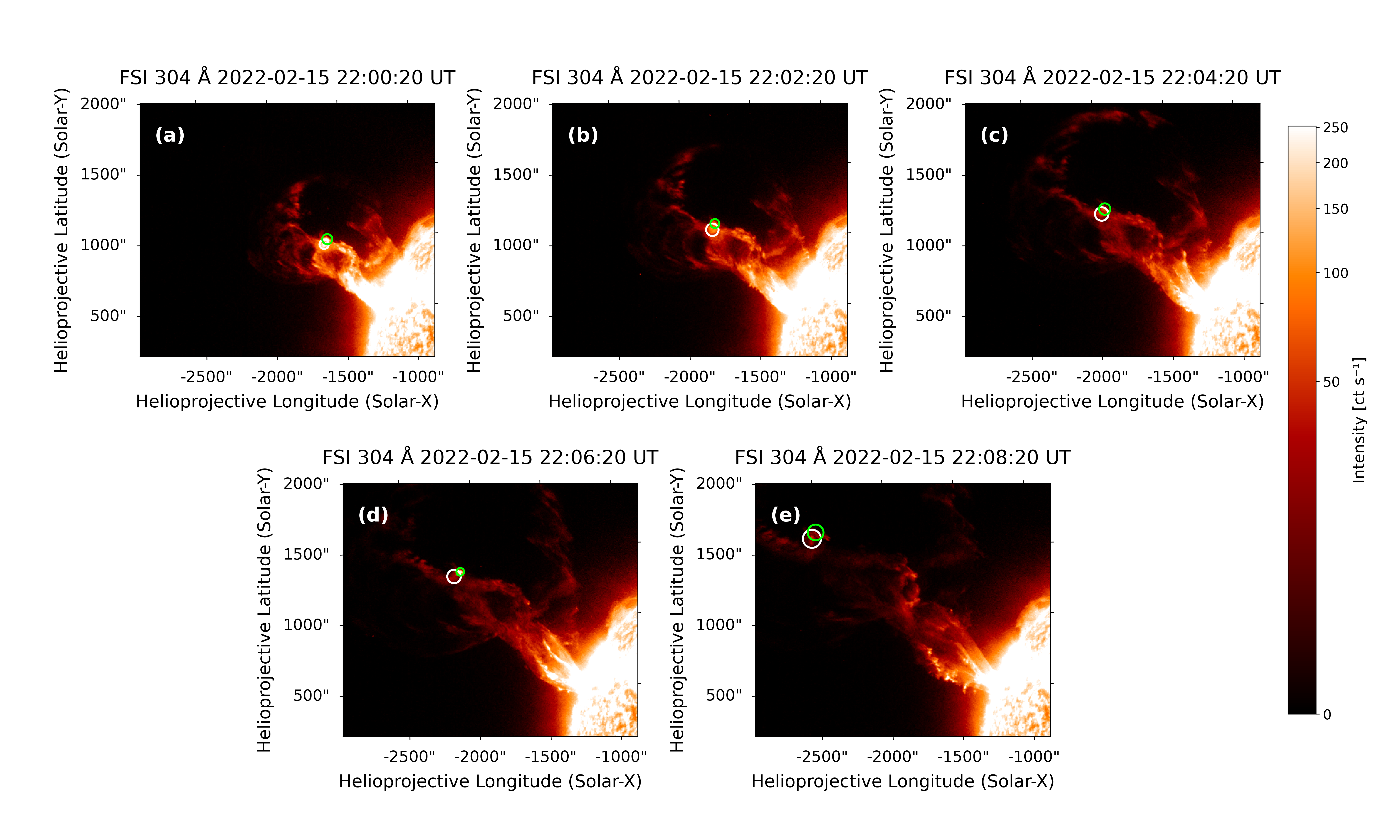}
    \caption{Evolution of the prominence eruption as seen by FSI in 304~\AA\ passband showing the two features being tracked.}
    \label{fig:track}
\end{figure*}
}
\afterpage{
\begin{figure*}
    \centering
\includegraphics[width=0.8\textwidth]{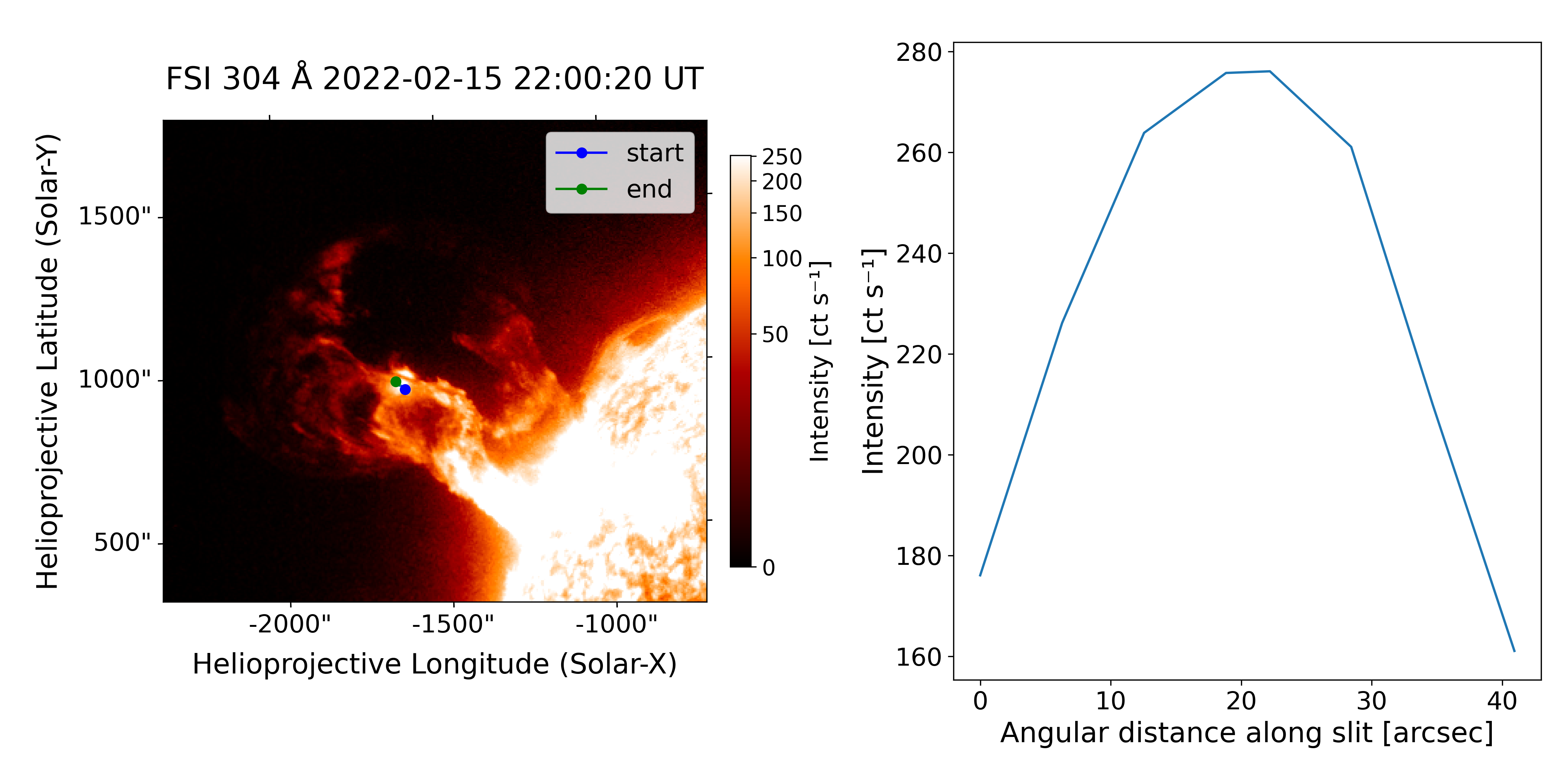}
    \caption{An example of the tracked feature in FSI 304~\AA\ passband at 22:00 UT, February 15, 2022. The right image is the intensity profile along the cut in the left image.}
    \label{fig:Feature_track}
\end{figure*}
}

To find the actual variation in intensity, we used images recorded at 21:00~UT before the eruption and subtracted the background intensity from each pixel in the images taken during the prominence eruption. Table~\ref{table:171_195} summarises the intensity recorded for two filters before and after subtracting the background. We obtained the minimum (maximum) intensity ratio of 0.30 (0.64) for five features. We plotted these minimum and maximum intensity ratio values as a red dashed lines in Fig.~\ref{fig:Response} to narrow down the range of ratio and temperature values (log~\textit{T} [K] = 5.20\textasciitilde5.40 (0.16\textasciitilde0.25~MK), and 6.12\textasciitilde6.18 (1.31\textasciitilde1.51~MK) and a few more values in the higher temperature range) we should consider. The temperature range of log~\textit{T} [K] = 5.20\textasciitilde5.40 (0.16\textasciitilde0.25~MK) is more appropriate for the features along the eruptive prominence we are studying in this paper.



\section{Results}
\label{Sec: compare}
\subsection{Intensity change of 304~\AA\ line with altitude}

We have explored how a feature's intensity varies with altitude during the eruption process in this event. We have tracked two bright regions (indicated by grey and green circles in Fig.~\ref{fig:track}) in FSI 304~\AA\ passband. The cadence of the FSI data in the 304~\AA\ passband was around 2 minutes. The 304~\AA\ passband of EUVI has a cadence of around 3 minutes. Therefore, we use FSI data to track features. The radius of the circles is proportional to the size of the visual edges of the features: if the circle becomes smaller, the feature becomes more concentrated in plasma density; if the circle becomes larger, the feature becomes more dispersed. After tracking the two features, we trace a diagonal cut through the bright region and obtain an intensity profile along that cut.  Figure~\ref{fig:Feature_track} shows an example of the tracked feature in FSI 304~\AA\ passband at 22:00 UT, February 15, 2022. The left panel shows the start and the end of a diagonal cut of a feature;  the right panel shows the intensity profile along the cut from the start to the end. To reduce the uncertainty in the diagonal cut of the features, we calculate the mean intensity along the diagonal line after removing 15\% of the data on the left side of the intensity profile and 15\% from the right side.

\begin{figure*}
    \centering
\includegraphics[width=\textwidth]{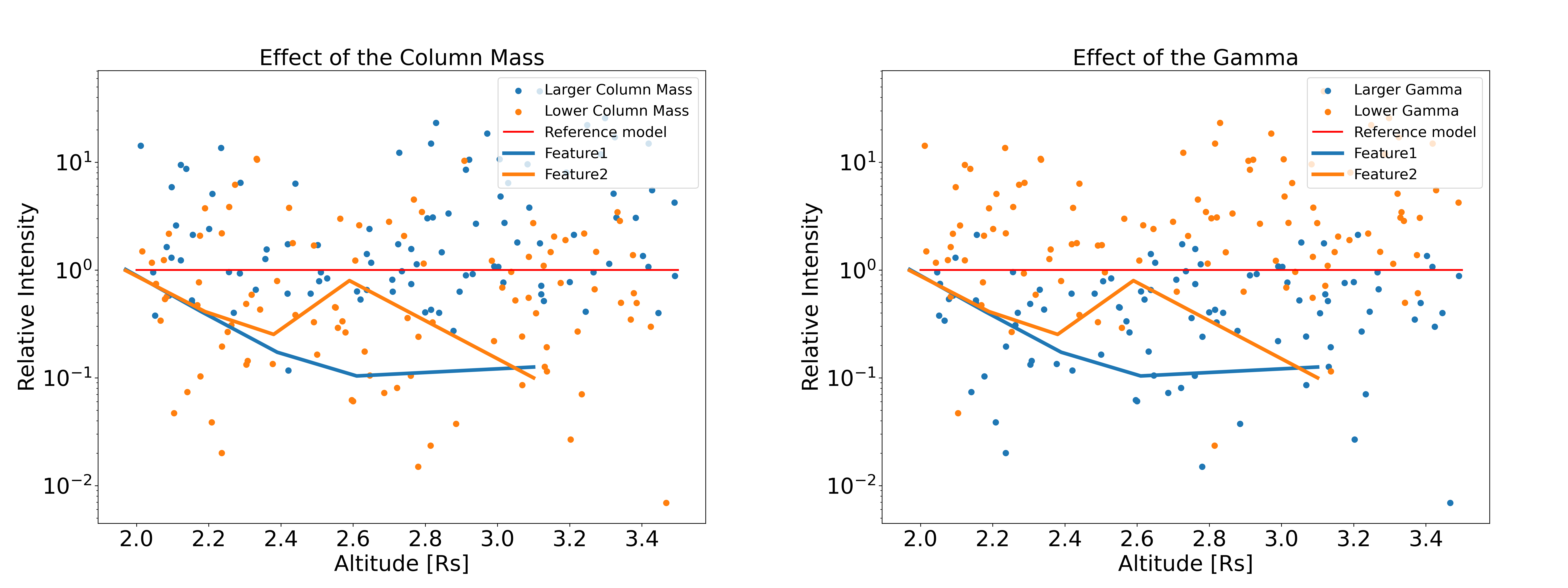}
    \caption{The change in the relative intensity for features 1 and 2 with altitude. The points correspond to 200 random models. The orange and blue lines show intensity variation with altitude for the two features tracked over time by the observation. The relative intensity of the two features is normalised by their intensity at $R_S=1.97$. The red lines are connected by reference models that have fixed values of parameters except altitude. The values of parameters have been set as the median of their ranges. The relative intensity of random models and reference models is normalised by the reference model at $R_S=2$. }
    \label{Effect_altitude}
\end{figure*}

In Fig.~\ref{Effect_altitude}, we can see that the intensity of the two features decreases quickly until they reach an altitude of 2.4~$R_s$. Above that, Feature 1 changes slowly, whereas the intensity of Feature 2 suddenly increases at around 2.6~$R_s$. The high intensity value occurs at 22:06:20~UT in Fig.~\ref{fig:track}. We can see that the circle for Feature 2 is quite small compared with other panels, which means Feature 2 is more concentrated at this time. This might be due to the effect of plasma motions. The radial velocity of different regions within the feature may vary, leading to changes in plasma density over time. As a result, the feature may appear more concentrated or diffuse at different times. 

There could also be material changes inside and outside the circles. These material changes would affect the intensity we observe in Fig.~\ref{Effect_altitude}. When some material moves into the circle, the intensity increase; when some material moves out of the circle, the intensity decreases. However, we think this effect should be negligible because the two features are relatively stable.

We generate 200 random PCTR models to compare with the observation, as shown in Table~\ref{table:parameter_ranges}. The choice of the altitude depends on data, in the range $2 \sim 3.5~R_S$. In section~\ref{Sect4:filter-ratio}, we refined the temperature range as $\log _{10}\left(T\right) [K]$ = 5.20\textasciitilde5.40 (0.16\textasciitilde0.25~MK) using the filter ratio technique. We constrained the temperature range using observations of the 171~\AA\ and 195~\AA\ lines, which are emitted primarily from a thin PCTR region. Therefore, we adopt this range as the surface temperature range. Our derived temperature range is roughly consistent with the findings of \cite{parenti2012nature}, who showed that the 171~\AA\ band in prominences is dominated by \ion{Fe}{9} emission from plasma at temperatures $>4\times10^{5}$~K ($\log _{10}\left(T\right) [K] > 5.6$), and that such emission arises from the PCTR.

We obtain the range of radial velocity by tracking several features, and we identify the range of altitude from the image. The range of radial velocity we obtained is $1180~\text{km}~\text{s}^{-1} \sim 2200~\text{km}~\text{s}^{-1}$, which is similar to \cite{mierla_2022_prominence}. The uncertainty of the result is about $120~\text{km}~\text{s}^{-1}$, which comes from the uncertainty of the location of features. The range of altitude we obtained is $2.0~R_S \sim 3.5~R_S$. For other parameters, we choose the usual ranges that are expected in a prominence event from \cite{labrosse2012plasma}.






\begin{table}
    \centering
    \caption{\centering The parameter ranges for different variables.}
    \resizebox{0.7\linewidth}{!}{
    \begin{tabular}{|l|l|}
    \hline
    \textbf{Parameter} & \textbf{Range} \\ \hline
    Central Temperature & $8000~\text{K} \sim 15000~\text{K}$ \\
    Surface Temperature & $158500~\text{K} \sim 251000~\text{K}$ \\
    Central Pressure & $0.001~\text{dyn}~\text{cm}^{-2} \sim 1.0~\text{dyn}~\text{cm}^{-2}$ \\
    Surface Pressure & $0.001~\text{dyn}~\text{cm}^{-2} \sim 0.1~\text{dyn}~\text{cm}^{-2}$ \\
    Column Mass & $1.0 \times 10^{-7}~\text{g}~\text{cm}^{-2} \sim 1.0 \times 10^{-5}~\text{g}~\text{cm}^{-2}$ \\
    $\gamma$ & $2 \sim 10$ \\
    Altitude & $1392680~\text{km} \sim 2437190~\text{km}$ \\
    Radial Velocity & $500~\text{km}~\text{s}^{-1} \sim 2500~\text{km}~\text{s}^{-1}$ \\
    Microturbulent Velocity & $5~\text{km}~\text{s}^{-1}$ \\ \hline
    \end{tabular}
    }
    \label{table:parameter_ranges}
\end{table}

\begin{table}
    \centering
    \caption{\centering The reference model parameters.}
    \resizebox{0.5\linewidth}{!}{
    \begin{tabular}{|l|l|}
    \hline
    \textbf{Parameter} & \textbf{Value} \\ \hline
    Central Temperature & $11500~\text{K}$ \\
    Surface Temperature & $204750~\text{K}$ \\
    Central Pressure & $0.50~\text{dyn}~\text{cm}^{-2}$ \\
    Surface Pressure & $0.05~\text{dyn}~\text{cm}^{-2}$ \\
    Column Mass & $5.05 \times 10^{-6}~\text{g}~\text{cm}^{-2}$ \\
    $\gamma$ & $6$ \\
    Radial Velocity & $1500~\text{km}~\text{s}^{-1}$ \\
    Microturbulent Velocity & $5~\text{km}~\text{s}^{-1}$ \\ \hline
    \end{tabular}
    }
    \label{table:reference_model}
\end{table}

In Fig.~\ref{Effect_altitude}, we study the change in relative intensity with altitude. To better compare with observations, we define a series of reference models and use relative intensity to track the trend of intensity changes through altitude. The reference models have values of parameters that are set to the median of their ranges, except for the altitude, as shown in Table~\ref{table:reference_model}. The reference models of different altitudes are connected as a red line in Figure~\ref{Effect_altitude}.

We can see that the relative intensity nearly has no change with altitude for reference models that have fixed values of parameters except altitude (the red line). Therefore, the change in intensity with altitude in the observations is likely to be caused by the change in other parameters. In Fig.~\ref{Effect_altitude}, among the brighter models, 64~\% of them have larger column mass, 85~\% of them have lower gamma; among the dimmer models, 63~\% of them have lower column mass, 78~\% of them have larger gamma. This suggests that column mass and gamma play important roles in the change of intensity with altitude, which could explain the intensity changes of the two features (orange and blue lines) in Fig.~\ref{Effect_altitude}.

\subsection{Analysis of parameters' effect by parallel coordinate plots}
\label{sec:effect}
We use the method of the parallel coordinate plot to analyse the prominence plasma parameters' effects on quantities related to the formation of the 304~\AA\ line. A parallel coordinate plot is a method for visualising relationships in multivariate data. Parallel coordinate plots have proven effective for understanding complex spatiotemporal datasets, particularly when spatial, temporal, and attribute-based relationships are crucial \citep{edsall2003parallel}. With this method, we can see a parameter's effect more easily without setting other parameters into a specific range. We can also simultaneously display and compare values and relationships of multiple variables. 

This approach has been successfully applied in our previous studies to investigate the formation of the hydrogen Lyman lines in solar prominences. In \cite{zhang2026non}, we used parallel coordinate plots to analyse the effects of different parameters of Lyman $\beta$ and Lyman $\gamma$ lines, such as temperature, pressure, column mass, and slab thickness. This provided a framework to interpret the SPICE observations presented in \cite{zhang2026analysis}. Those studies demonstrated that the method is particularly useful for identifying dominant parameters and exploring relations among them. Building on this earlier work, we apply the same methodology to the \ion{He}{2} 304~\AA\ line, exploring helium-line formation processes in prominence plasmas.

To analyse the FSI observations of this event, we ran 200 random PCTR models. The event has observational constraints for surface temperature (from filter-ratio analysis), altitude, and radial velocity. 

\begin{figure*}
    \centering
    \includegraphics[width=0.95\textwidth]{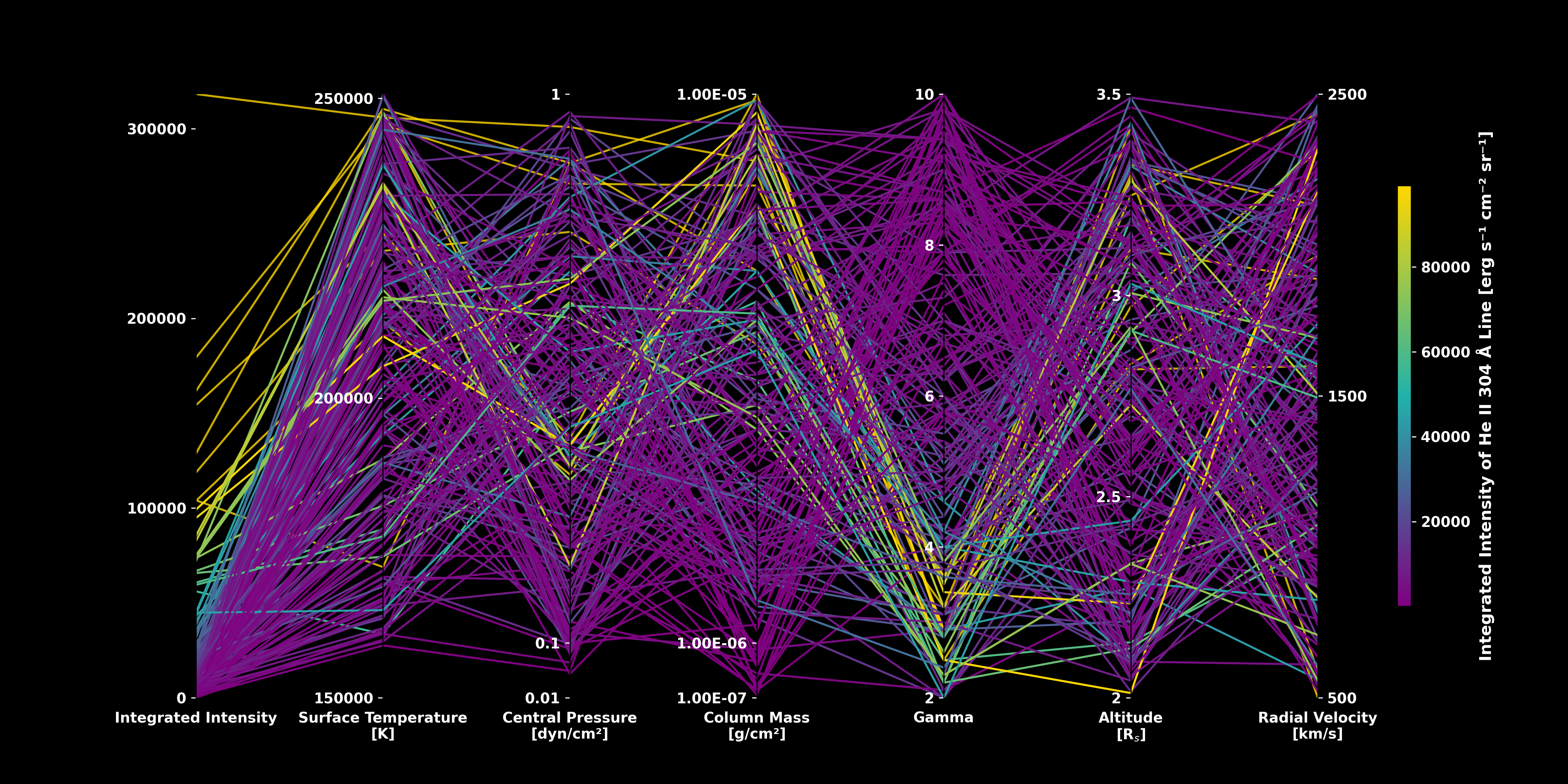}
    \caption{The parallel coordinate plot of 6 parameters with the integrated intensity of 304~\AA\ line.}
    \label{fig:II304}
\end{figure*}

In a parallel coordinate plot like Fig.~\ref{fig:II304}, if we see a diversity of colours on different value ranges when we focus on one parameter's axis, this means the parameter does not have much effect; on the other hand, if we see a certain colour concentrating on a certain value range, this means the parameter has some effect. The effect is more significant when the the range of values over which a certain colour concentrates is narrow. It is also possible to use correlation coefficients to quantify these relationships. We have noted more obvious effects on a parallel coordinate plot when the corresponding correlation coefficient is larger than 0.3. Hence, when a correlation coefficient between, e.g., one of the plasma parameters and one characteristic of the 304~\AA\ line is larger than 0.3, we will consider that parameter as an important parameter in this case.


We explore what parameters are important for the integrated intensity and the optical thickness of the 304 \AA\ line. In Fig.~\ref{fig:II304}, we can see that the most important parameters of the integrated intensity of the 304 \AA\ line are column mass and gamma, which have correlation coefficients with the integrated intensity of 0.36 and -0.51, respectively. 

\begin{figure*}
    \centering
    \includegraphics[width=0.95\textwidth]{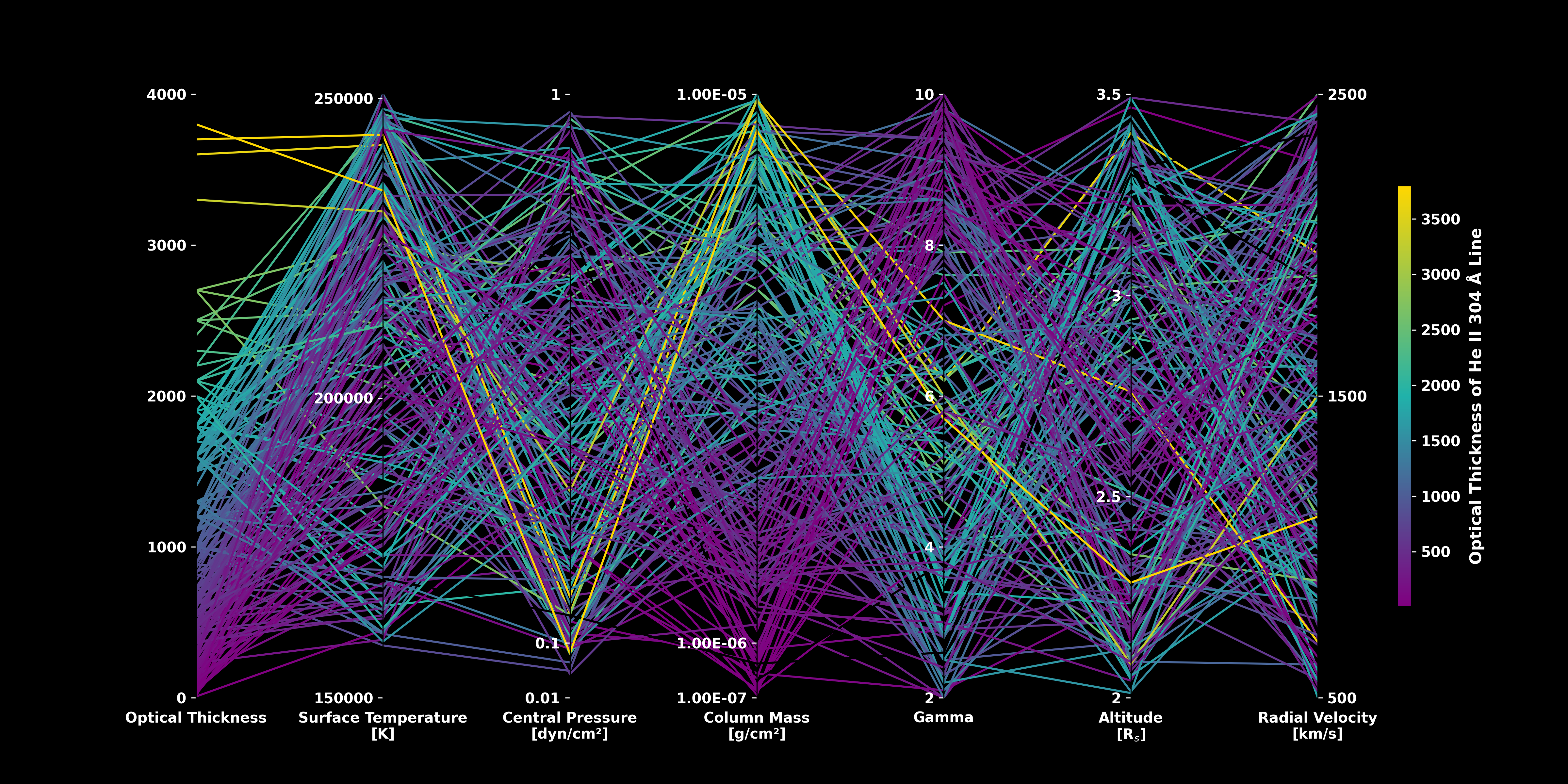}
    \caption{The parallel coordinate plot of 6 parameters with the optical thickness of 304~\AA\ line.}
    \label{fig:OT304}
\end{figure*}

In Fig.~\ref{fig:OT304}, we can see that the most important parameter of the optical thickness of the 304 \AA\ line is column mass, which has a correlation coefficient with optical thickness of 0.75. 
Correlation coefficients of key parameters for the \ion{He}{2} 304~\AA\ line are listed in Table~\ref{tab:heii304_corr}.

\begin{figure*}
    \centering
    \includegraphics[width=0.95\textwidth]{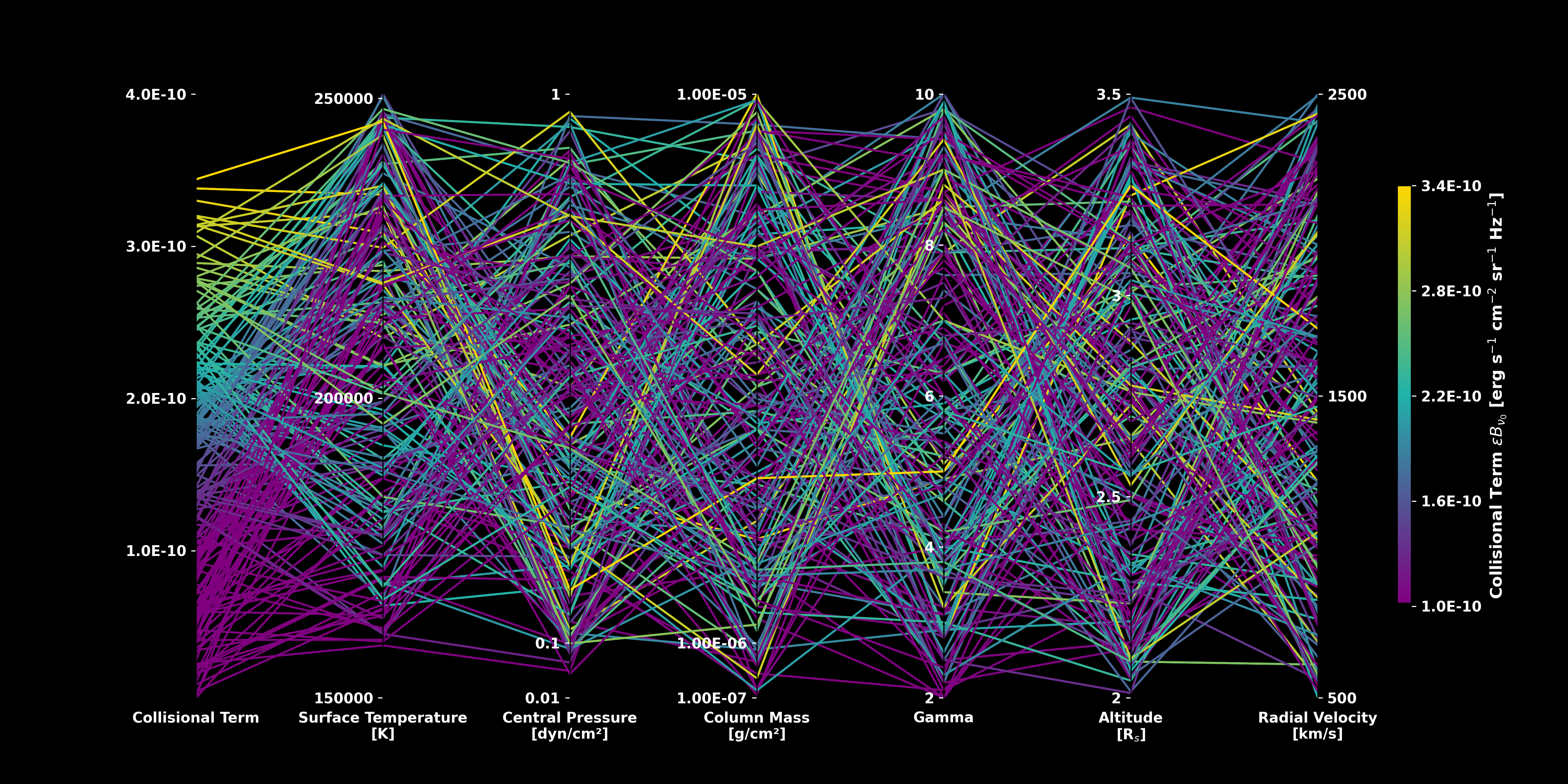}
    \caption{The parallel coordinate plot of 6 parameters with the collisional term of 304~\AA\ line. }
    \label{fig:CT304}
\end{figure*}
\begin{figure*}
    \centering
    \includegraphics[width=0.95\textwidth]{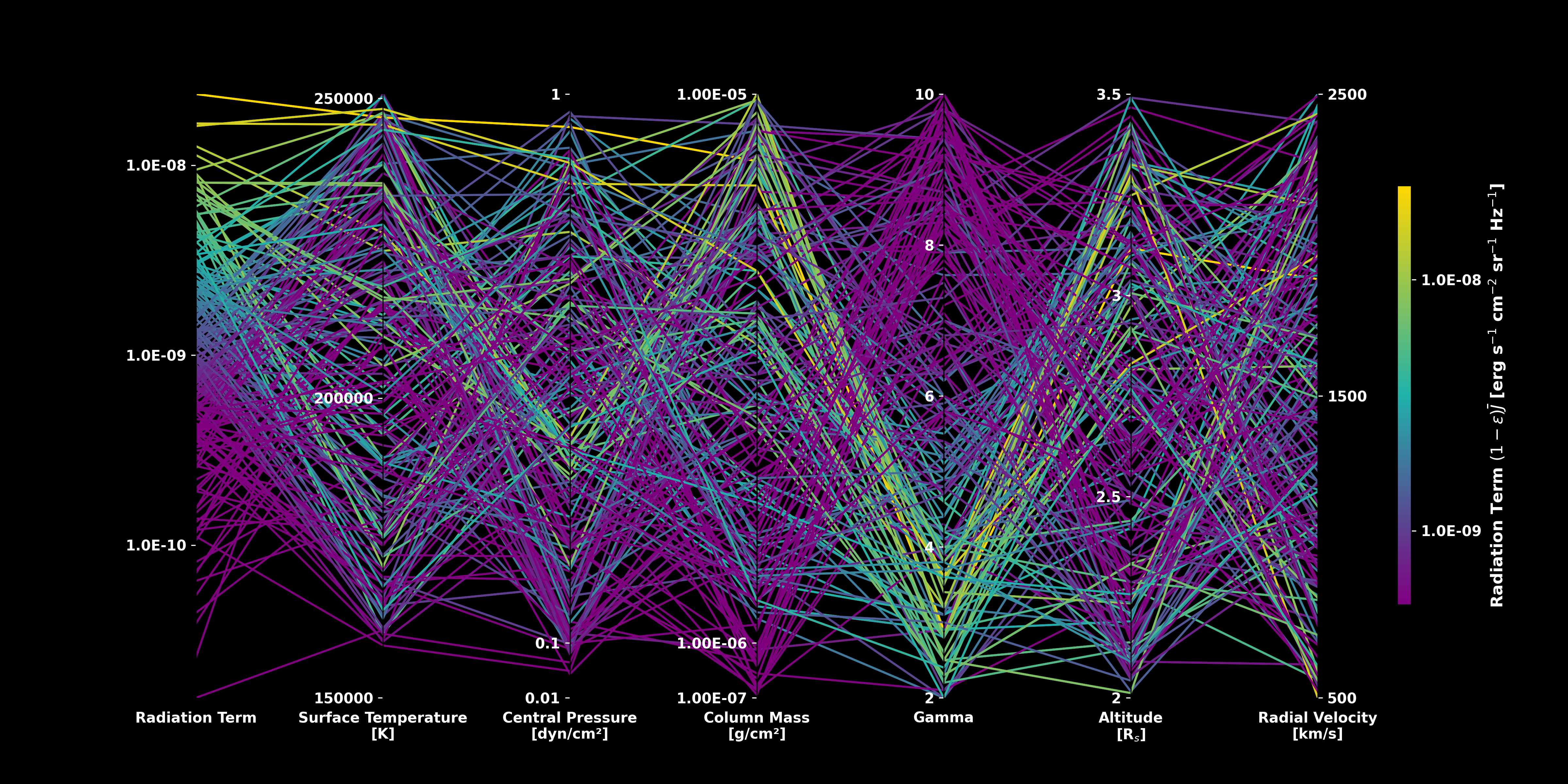}
    \caption{The parallel coordinate plot of 6 parameters with the radiative term of 304~\AA\ line. }
    \label{fig:RT304}
\end{figure*}

\begin{deluxetable}{lcc}
\tablecaption{Correlation coefficients for \ion{He}{2} 304~\AA\ line properties\label{tab:heii304_corr}}
\tablehead{
\colhead{Property} & \colhead{Key Parameters} & \colhead{Correlation Coefficient}
}
\startdata
\multirow{2}{*}{Integrated Intensity} & Column mass & 0.36 \\
                                       & Gamma      & -0.51 \\
\hline
\multirow{2}{*}{radiative term}        & Column mass & 0.36 \\
                                       & Gamma      & -0.55 \\
\hline
Optical Thickness                      & Column mass & 0.75 \\
\enddata

\tablecomments{The parameters that don't appear in the table correspond to correlation coefficients less than 0.3. There is no 'Collisional Term' in this table because no parameter has a correlation coefficient larger than 0.3 with the collisional term.}
\end{deluxetable}
To study how collisional excitation and radiative processes contribute to the \ion{He}{2} 304~\AA\ line formation, we generate Figs.~\ref{fig:CT304} and~\ref{fig:RT304} to study the contributions of the collisional term and radiative term of the 304 \AA\ line. We use the collisional term and radiative term from the source function given by equation \ref{Eq: source} \citep{labrosse2006lyman}. 

\begin{eqnarray}
S & = & (1 - \epsilon) \bar{J} + \epsilon B_{\nu_0}
\label{Eq: source}
\end{eqnarray}
$\epsilon$ is the probability of a photon destruction. $\bar{J}$ is the mean intensity of the radiation field. $B_{\nu_0}$ is the Planck source function. The collisional term is $\epsilon B_{\nu_0}$ and the radiative term is $(1 - \epsilon) \bar{J}$. We consider the values of the collisional and radiative terms taken at the surface of the prominence, which is the position where the contribution function of the \ion{He}{2} 304~\AA\ line peaks. 

In Fig.~\ref{fig:CT304}, we can see that none of the six parameters considered in this study has a significant effect on the collisional term. All 200 random models lead to collisional terms that are lower than $4.0 \times 10^{-10}\,\mathrm{erg\,s^{-1}\,cm^{-2}\,sr^{-1}\,Hz^{-1}}$.  
Figure~\ref{fig:RT304} shows that the most important parameters to determine the radiative term in the formation of the 304 \AA\ line are column mass and gamma, which have correlation coefficients (between parameters and the radiative term) of 0.36 and -0.55. Among the 200 random models, most of them have radiative terms from $1.0 \times 10^{-10}\,\mathrm{erg\,s^{-1}\,cm^{-2}\,sr^{-1}\,Hz^{-1}}$ to $1.0 \times 10^{-8}\,\mathrm{erg\,s^{-1}\,cm^{-2}\,sr^{-1}\,Hz^{-1}}$. This result shows that the radiative term is dominant in the formation of the \ion{He}{2} 304~\AA\ line for the range of model parameters considered here. 

\subsection{Discussion}
\label{Sect5:discussion}
\subsubsection{Effect of model parameters on the \ion{He}{2} 304~\AA\ line}

We have analysed the most important parameters or the February 15, 2022, event in relation to the formation of the \ion{He}{2} 304~\AA\ line. 
%
%
%
To summarise, we found that the integrated intensity of the \ion{He}{2} 304~\AA\ line is mainly determined by the column mass and by gamma (corresponding to the temperature gradient in the PCTR). The optical thickness of the line is mostly affected by the column mass. The radiative term in the source function is mostly affected by column mass and gamma. The model parameters did not show a significant effect on the collisional term in the source function.


The column mass affects the integrated intensity, the radiative term and the optical thickness. 
The column mass affects radiation rates, which enter the radiative term of the expression of the source function (Eq. \ref{Eq: source}).  $\bar{J}$ is the mean intensity of the radiation field. When we vary the column mass while keeping the other parameters fixed, the density of particles does not change much. Thus, the total intensity of the radiation field positively correlated with column mass. Through the source function, the column mass can indirectly affect the integrated intensity. 

The value of gamma (which corresponds to the steepness and extent of the PCTR) affects both the integrated intensity and the radiative term. When gamma is high, it means that the temperature changes more quickly in the PCTR, which means the high-temperature part of the prominence is less extended. This directly impacts the integrated intensity. The temperature structure affects the amount of plasma that efficiently participates in radiative processes. Therefore, gamma also affects the radiative term.


The surface temperature, central pressure, altitude and radial velocity have no significant effects compared with other parameters in the parameters' ranges we have refined.


In this study, we have assumed that the observed emission in the 304~\AA\ passband is entirely due to \ion{He}{2}. However, \ion{Si}{11} at 303.32~\AA\ which forms at $\sim1.5$~MK, as well as other ions such as \ion{O}{5}, \ion{Ca}{18} and \ion{Fe}{18} also contribute to the observed 304~\AA\ line. In off‑limb observations, the emission of those lines can be comparable to the \ion{He}{2} 304~\AA\ signal, especially over the diffuse hot corona \citep{antolin2024decomposing}. \cite{antolin2024decomposing} presented several methods to separate the cool and hot components within the AIA 304~\AA\ passband, showing that the hot contribution can be significant and that its removal improves the contrast of cool structures such as coronal rain and prominences. Although our present analysis does not explicitly correct for this hot contamination, its presence could affect the absolute intensity levels we derive. Nevertheless, our study focuses on the relative behaviour of the \ion{He}{2} 304~\AA\ line with altitude and the roles of collisional and radiative processes. The main conclusions regarding the importance of central pressure, column mass and temperature gradient steepness would not change if we took the contributions of these other lines into account. However, future analysis of 304~\AA\ line study could benefit from applying similar decomposition techniques to isolate the pure \ion{He}{2} 304~\AA\ component, especially when studying off‑limb eruptive prominences.

\subsubsection{Formation of the \ion{He}{2} 304~\AA\ line}

Our findings can be compared with the earlier work of \citet{labrosse2012plasma}, who investigated the formation of the \ion{He}{2} 304~\AA\ line in eruptive prominences using SDO/AIA observations and non-LTE radiative transfer modelling. \citet{labrosse2012plasma} demonstrated that the computed intensity of the 304~\AA\ line may either decrease or increase with radial velocity depending on the local plasma conditions, particularly the column mass and temperature of the prominence plasma. They highlighted that an increase in column mass during an eruption could counteract the Doppler dimming effect and even lead to an overall brightening of the line with velocity. 

In the present study, we extend this analysis by systematically exploring the effects on the parameters using a set of 200 random PCTR models and parallel coordinate plots to visualise multivariate relationships. We confirm that column mass plays a critical role in determining the integrated intensity, the radiative term, and the optical thickness of the \ion{He}{2} 304~\AA\ line. Furthermore, we identify additional key parameters, central pressure and steepness of the temperature gradient ($\gamma$), which significantly influence the line formation in this event. 

In our study, radial velocity is not an important parameter, which is different from \citet{labrosse2012plasma}. This is because the  radial velocity values in our study are quite large in comparison to \citet{labrosse2012plasma}, which leads the incident photons at 304~\AA\ to be out of resonance with the \ion{He}{2} ions. 
\cite{mierla_2022_prominence} conjecture that the 304~\AA\ line in this eruptive prominence is due to collisional excitation rather than to resonant scattering. 
The high altitude of the prominence will indeed lead to a strong dilution of the incident radiation, and the large radial velocity limits the efficiency of resonant scattering of incident photons by \ion{He}{2} ions due to the Doppler dimming effect. It is therefore reasonable to conclude that resonant scattering of the incident radiation cannot explain the emission observed.
However, our study shows that, in the source function, the collisional term is negligible and that radiative processes dominate the formation of the \ion{He}{2} 304~\AA\ line over the range of model parameters considered here to interpret EUI eruptive prominence observations. 
We conclude that the emission in the \ion{He}{2} 304~\AA\ line in this fast-moving, high-altitude eruptive prominence is radiatively controlled, namely via radiative recombination of  \ion{He}{3} followed by cascades populating the \ion{He}{2}  $n=2$ state followed eventually by spontaneous emission in 304~\AA. This is supported by the fact that at the high temperatures we have identified in the region of formation of the line (in the PCTR close to the prominence surface), helium is mostly fully ionised -- there aren't enough \ion{He}{2} ions in the ground state to be collisionally excited.

\section{Conclusions}
\label{Sect6:conclusion}

In this study, we analysed the eruption of a solar prominence observed by the EUI instrument aboard Solar Orbiter in the \ion{He}{2} 304~\AA\ line, focusing on various physical parameters such as temperature, radial velocity, and altitude. 
A filter-ratio analysis using the 171~\AA\  and 195~\AA\ channels of STEREO/EUVI-A allows us to constrain the temperature of the eruptive prominence around log~\textit{T} [K] = 5.20\textasciitilde5.40. We also constrain the radial velocity in the range 1180\textasciitilde2200~km~s$^{-1}$ by tracking different features in the eruption. By analysing various parameters and generating 200 random non-LTE PCTR models, we investigated how these parameters influence the formation of the \ion{He}{2} 304~\AA\ line.

We compared our results with \citet{labrosse2012plasma}. We confirm that column mass is an important factor in the formation of the He~II 304~\AA\ line. The effect of radial velocity is not visible in our study, which is due to the high radial velocity values we observed. 
Our results suggest that column mass and the steepness of the temperature profile ($\gamma$) are key factors affecting the \ion{He}{2} 304~\AA\ line's formation in this event, as indicated by their strong correlations with integrated intensity, radiative term and optical thickness. The study also revealed the impact of column mass and gamma on intensity variation with altitude during the eruption, which displays how physical conditions within the prominence change with altitude.  

Our study also shows that at temperatures of a few $10^5$~K, \ion{He}{2} 304~\AA\  emission in a high-altitude, outward-moving prominence is no longer produced by resonant scattering of solar radiation. Instead, it is produced predominantly by radiative recombination of \ion{He}{3} followed by cascades into the $n=2$ level. Radiative rates dominate over collisional destruction rates. Thus, the emission remains radiatively dominated, but the dominant radiative process has shifted from scattering of the incident radiation (in the case of low-lying, relatively quiescent or slowly moving prominences) to photoionisation followed by recombination.
This is different from the conjecture in \cite{mierla_2022_prominence}.

While this analysis provides significant insights into prominence dynamics and the \ion{He}{2} 304~\AA\ line's formation, it is limited by the lack of calibrated intensities. In future work, we will combine the work of \cite{zhang2026analysis}, \cite{zhang2026non}, Zhang et al.~(2026c, submitted), and this paper to analyse hydrogen and helium lines observed simultaneously in the same event, e.g. by SPICE and EUI.

\section*{Acknowledgements}

YZ is supported by the China Scholarship Council (No. 202206120056). NL and SM acknowledge support from UK Research and Innovation's Science and Technology Facilities Council under grant award numbers ST/T000422/1 and ST/X000990/1. We would like to thank the teams involved in the development and operation of the EUI and EUVI instruments and data release. We are also grateful to the SunPy team for providing the tools used in our data analysis. We also thank Dr Iain Hannah for useful discussions, and thank Susanna Parenti and Therese Kucera for their useful comments. We thank Dr. Markus Aschwanden, Dr. Nariaki Nitta, Dr. Peter Young, Dr. Giulio Del Zanna, and Dr. Roger Dufresne for discussions on the STEREO EUVI temperature responses. CHIANTI is a collaborative project involving George Mason University, the University of Michigan (USA), the University of Cambridge (UK) and NASA Goddard Space Flight Center (USA). 

In this paper, we used the SunPy open-source software package \citep{sunpy_community2020} and the software library (\url{https://zenodo.org/records/14919949}) to download EUI and EUVI-A data. 
We used SunPy version 5.0.0 for data analysis.
Solar Orbiter is a mission of international cooperation between ESA and NASA, operated by ESA.

\bibliography{sample701}{}
\bibliographystyle{aasjournalv7}



\end{document}